\documentclass[pdflatex,sn-mathphys-num]{sn-jnl}

\usepackage{graphicx}%
\usepackage{multirow}%
\usepackage{amsmath,amssymb,amsfonts}%
\usepackage{amsthm}%
\usepackage{mathrsfs}%
\usepackage[title]{appendix}%
\usepackage{xcolor}%
\usepackage{textcomp}%
\usepackage{manyfoot}%
\usepackage{booktabs}%
\usepackage{algorithm}%
\usepackage{algorithmicx}%
\usepackage{algpseudocode}%
\usepackage{listings}%
\usepackage{subcaption}

\newcounter{extdatafigure}
\renewcommand{\theextdatafigure}{\arabic{extdatafigure}}

\newenvironment{extdatafigure}[1][]{
  \begin{figure}[#1]
  \refstepcounter{extdatafigure}%
}{
  \end{figure}
}

\newcommand{\extdatacaption}[1]{%
  \par\vskip\abovecaptionskip
  {\footnotesize\noindent\textbf{Extended Data Figure~\theextdatafigure.} #1\par}
}

\theoremstyle{thmstyleone}%
\theoremstyle{thmstyletwo}%

\theoremstyle{thmstylethree}%

\begin{document}

\title[Hashprice moderates the electricity demand response of Bitcoin miners]{Hashprice moderates the electricity demand response of Bitcoin miners}


\author[1,2]{\fnm{Subir} \sur{Majumder}}\email{subir-em@ieee.org}

\affil[1]{\orgdiv{Electrical and Computer Engineering Department}, \orgname{Texas A\&M University}, \orgaddress{\street{188 Bizzell St}, \city{College Station}, \postcode{77843}, \state{TX}, \country{USA}}}

\affil[2]{\orgdiv{School of Engineering and Applied Science}, \orgname{Harvard University}, \orgaddress{\street{150 Western Ave}, \city{Allston}, \postcode{02134}, \state{MA}, \country{USA}}}

\abstract{Large controllable loads, such as Bitcoin-mining facilities, are increasingly viewed as valuable sources of power-system flexibility, yet the conditions under which this flexibility is realized remain poorly understood. We examine this issue in the Texas power market, where large loads face both wholesale electricity prices and incentives created by coincident-peak-based transmission charges. We find that mining load declines as costs rise across both channels, and this response is moderated by hashprice, a measure of expected revenue for Bitcoin miners. When hashprice is higher, mining load is less responsive to electricity-sector costs. This pattern is consistent with aggregate mining load arising from heterogeneous devices operated around distinct breakeven points. The wholesale-price response illustrates this mechanism most clearly. Mining load remains largely online at low electricity prices but begins to decline once prices exceed an implied curtailment threshold, and higher hashprice shifts this threshold to higher wholesale prices. Bitcoin miners therefore respond to electricity-sector costs, but the available flexibility varies with revenue conditions in the crypto-financial sector. Treating such loads as stable demand-response resources may overstate their available flexibility.}


\maketitle

\section{Introduction}\label{sec1}

Modern power systems increasingly need flexible electricity demand. In many countries, aggregate short-run electricity demand has historically been only weakly to moderately responsive to electricity prices \cite{lijesen2007real, labandeira2017meta, burke2018price, zhu2018meta, hirth2024aggregate}. As renewable generation expands and peak-demand stress raises reliability concerns, loads that can adjust consumption when supply is scarce or system demand is high are becoming more valuable \cite{martinot2016grid, brunner2020future, IEA2025DemandFlexibility, suna2022assessment}. Large controllable loads are therefore increasingly viewed as potential sources of demand-side flexibility. Bitcoin mining is a prominent example of a large controllable load that can rapidly adjust electricity use in response to electricity-sector signals \cite{NERC_LargeLoads_2025, dlapiper2023_role_bitcoin_mining_renewables, PUCT2024_VirtualCurrencyMiningRegistration, gallant2024medicinehat, cpower2024blockfusion, blackhills2024blockchain}. Yet it remains unclear whether this flexibility is systematic enough to be treated as a predictable grid resource.

Bitcoin mining provides a useful empirical setting for studying this question because both sides of miners' short-run operating margin can be measured and can change over short time scales. Electricity is the dominant variable input cost for Bitcoin mining \cite{DEVRIES2018801}. Hashprice, which measures expected mining revenue per unit of computational power per unit time, summarizes the short-run value of continued operation \cite{hashrateindex2025bitcoinhashprice, neumueller2025cambridge}. Operationally, industry accounts and modeling assumptions suggest that mining devices are operated around breakeven points. A device can remain online when expected mining revenue, determined by hashprice and device efficiency, exceeds electricity costs, and is curtailed otherwise \cite{Bratcher2024, menati2023optimization, hajiaghapour2022approach, garratt2015entry}. Existing empirical work has linked Bitcoin prices, wholesale electricity-price volatility, and mining electricity use \cite{sapra2024uncovering, aye2023pricing}. Still, empirical evidence on Bitcoin miners' short-run curtailment behavior remains limited.

We examine Bitcoin miners' short-run curtailment behavior using the institutional setting of the Texas power market, where large loads can be exposed to two distinct electricity-sector cost channels: contemporaneous wholesale electricity prices and incentives created by coincident-peak-based transmission charges. Although we do not observe individual device-level operating decisions, the aggregate response is consistent with mining load being composed of heterogeneous devices operating around their respective breakeven points. Two patterns support this interpretation. First, Bitcoin-mining load declines as electricity-sector costs rise, but this response weakens when hashprice is higher. This state dependence appears across both cost channels. Second, the wholesale-price response reveals an implied curtailment threshold. At a given hashprice, mining load remains largely online at low electricity prices but begins to fall once prices rise sufficiently. Higher hashprice shifts this threshold to higher wholesale prices, consistent with stronger revenue conditions allowing marginal devices to remain profitable at higher electricity costs.

These findings imply that Bitcoin-mining flexibility cannot be treated as a fixed grid resource. Its availability depends partly on revenue conditions in the crypto-financial sector. For power-system operations, planning, and market design, large controllable loads should therefore be modeled as economically state-dependent sources of flexibility.

\section{Empirical design}\label{sec:design}

In the Texas power market, large loads can be exposed to two electricity-sector cost channels, wholesale electricity prices and coincident-peak-based transmission charges. Under coincident-peak charges, electricity consumption by large loads during system peaks determines their transmission charges for the following year \cite{baldick2018incentive, carmona2026coincident}. By consuming electricity during intervals that are likely to become coincident peaks, loads forgo the opportunity to reduce future transmission charges through curtailment. This creates an expected opportunity cost of electricity consumption. In Texas, coincident peaks are determined during the summer months, June through September \cite{PUCT1999_25_192}, so this expected opportunity cost is concentrated in those months. Large loads are also exposed to wholesale electricity prices throughout the year. We use this setting to examine how aggregate Bitcoin-mining load responds to wholesale electricity prices and coincident-peak incentives.

Because the mining-load data are aggregated at the load-zone level, our estimates characterize aggregate behavior across multiple Bitcoin-mining firms within each zone. We pool hourly observations across three Texas load zones.

\begin{figure}[!htbp]

\centering
\makebox[\linewidth][c]{%
\begin{minipage}{1.07\linewidth}
\centering

\begin{subfigure}{0.495\linewidth}
    \centering
    \includegraphics[width=\linewidth]{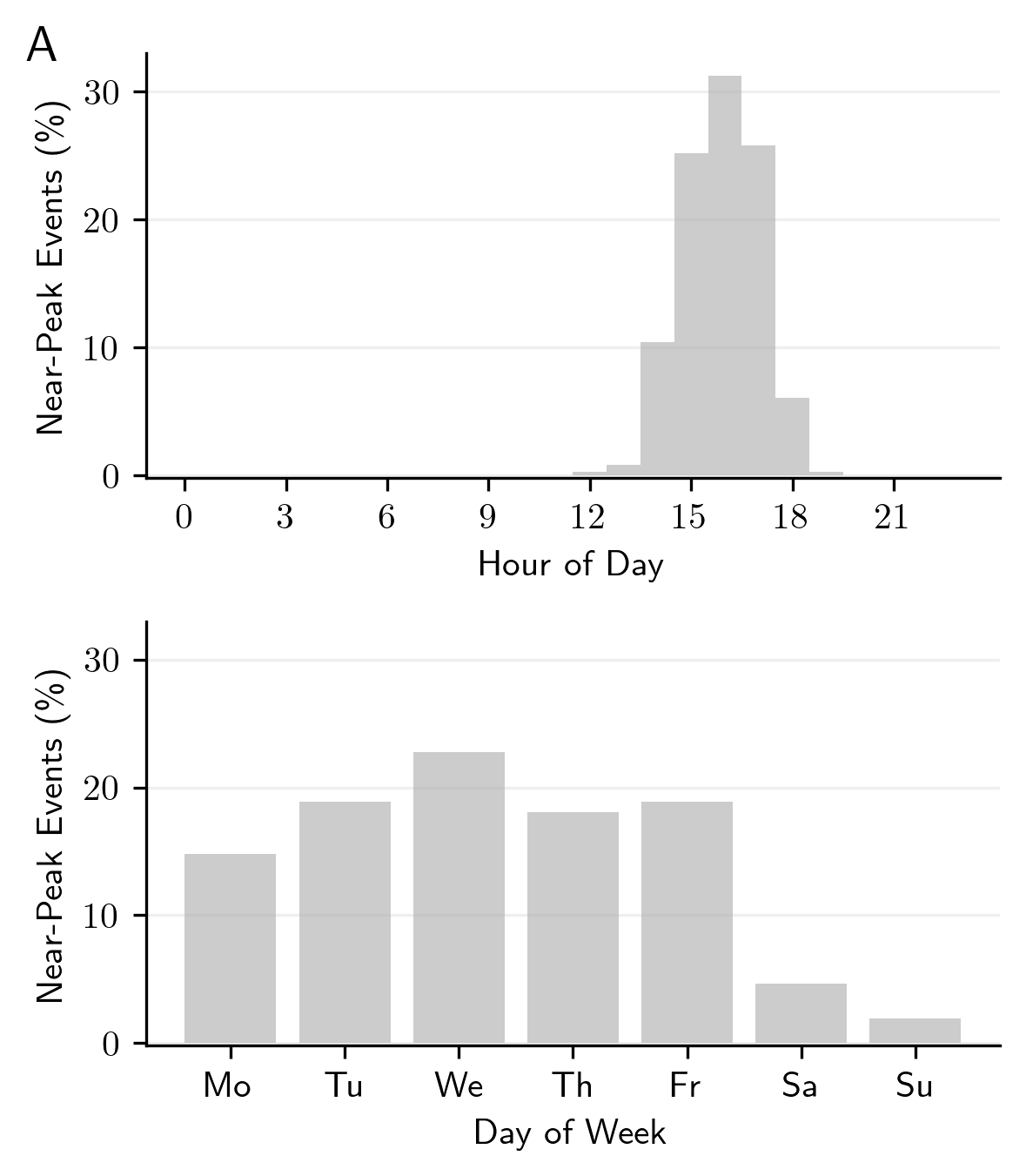}
    \caption*{}
    \label{fig2b}
\end{subfigure}
\hfill
\begin{subfigure}{0.495\linewidth}
    \centering
    \includegraphics[width=\linewidth]{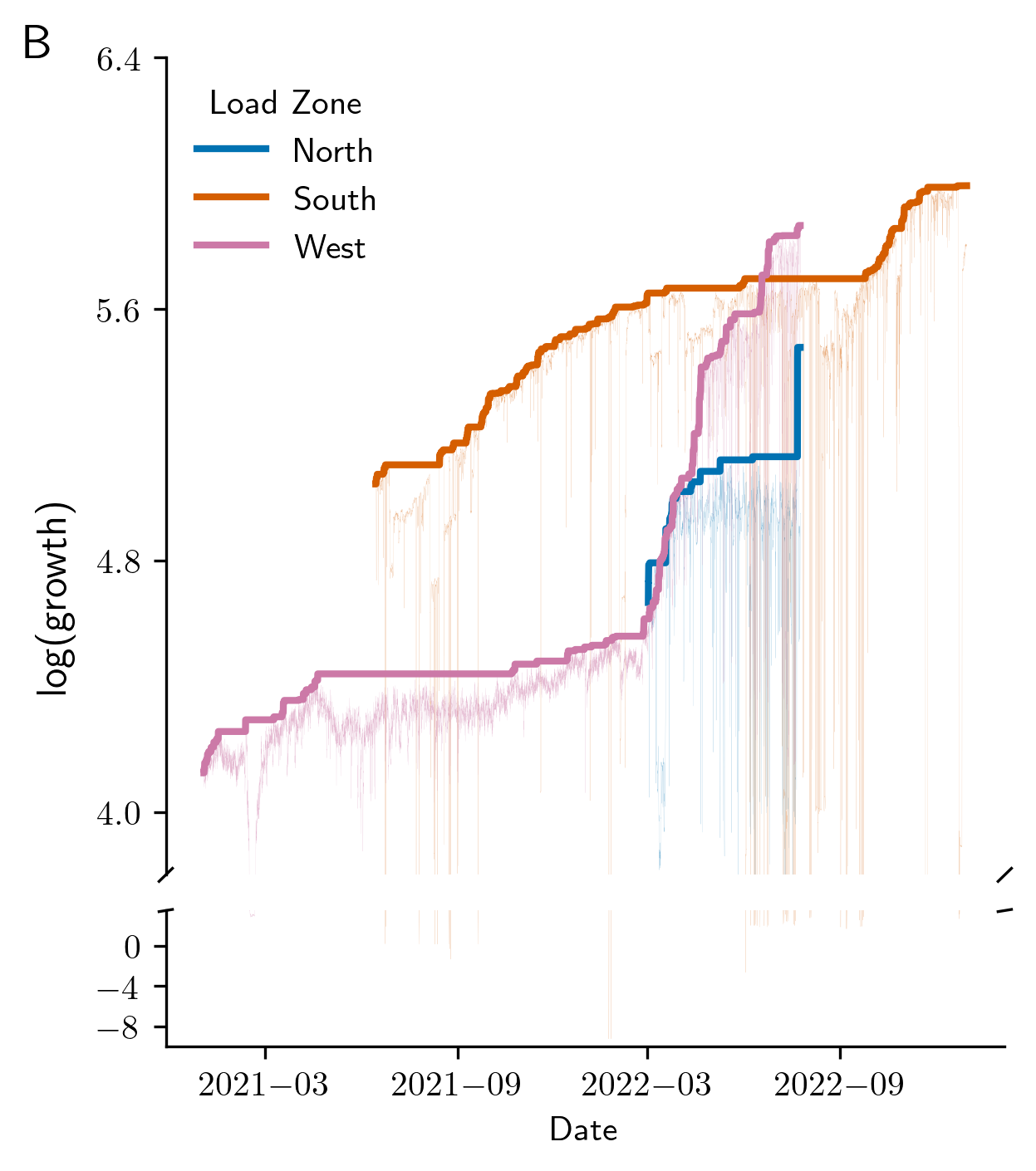}
    \caption*{}
    \label{fig2c}
\end{subfigure}

\vspace{-1.5em}

\begin{subfigure}{\linewidth}
    \centering
    \includegraphics[width=\linewidth]{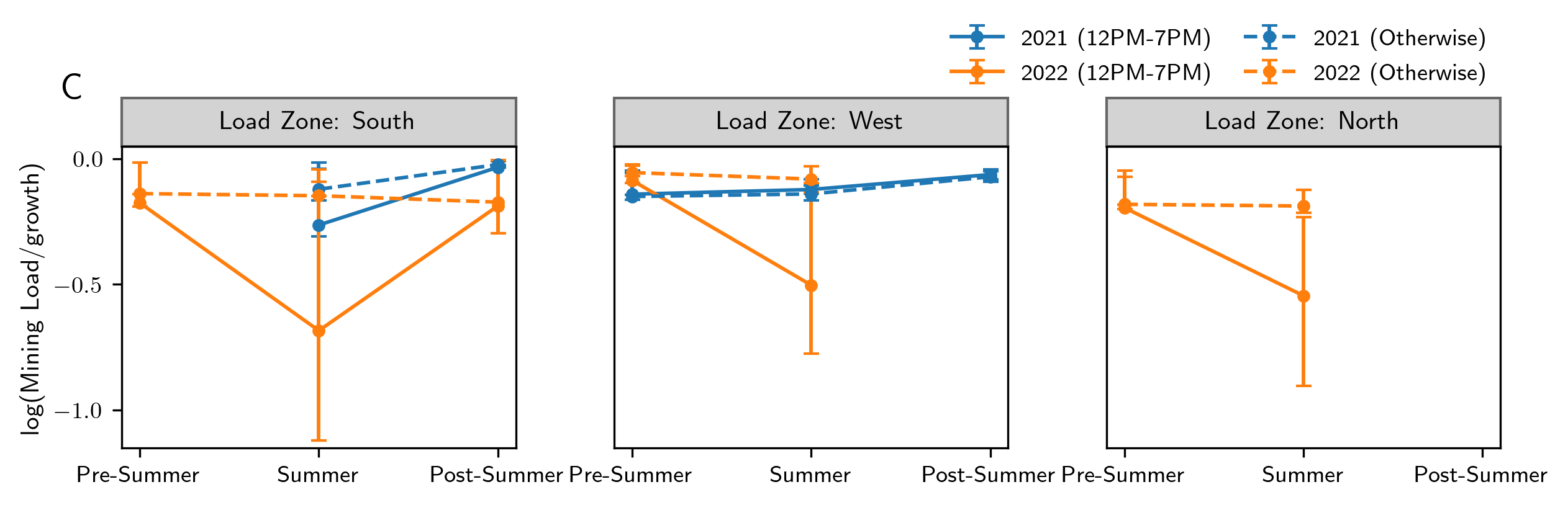}
    \caption*{}
    \label{fig2a}
\end{subfigure}

\end{minipage}%
}

\caption{Empirical definition of the summer--daytime window.
(A) Historical near-peak events in the Texas power grid. Events are classified as near-peak when June--September system demand is close to the monthly system peak. Their distribution by hour of day and day of week shows that near-peak conditions are concentrated between 12{:}00 and 19{:}00 Central Standard Time. This pattern motivates the summer--daytime (SDT) window.
(B) Load-zone-specific growth measures. Thick colored lines show the logged growth covariate for each load zone, constructed from prior mining-load observations outside the SDT window. Faint background lines show observed logged mining load.
(C) Growth-adjusted mining load by load zone, year, season, and SDT status. Points show average $\log(\text{Mining Load}/\text{growth})$ in pre-summer, summer, and post-summer seasons. Values are shown separately for 2021 and 2022 and for SDT-window hours versus all other hours. Vertical bars show the interquartile range within each year--zone--season--SDT-status cell. After adjustment for growth, non-SDT observations are relatively similar across observed seasons, whereas SDT-window observations show a larger summer reduction in 2022.}\label{fig2}
\end{figure}

Our empirical design has two parts. First, we estimate responsiveness to wholesale electricity prices while accounting for broad exposure to coincident-peak incentives using a summer--daytime (SDT) window. The SDT window is defined as June--September, 12{:}00--19{:}00 Central time, and captures periods when the Texas power grid is most likely to experience coincident peaks (Figure~\ref{fig2}A; see Methods). We interpret the SDT-window indicator as an intention-to-treat measure of exposure to coincident-peak incentives. This interpretation requires non-SDT observations to provide a valid counterfactual for SDT observations, conditional on controls and fixed effects \cite{roth2023s, angrist2009mostly}. To support this comparison, we construct load-zone-specific growth covariates that account for mining-load growth across zones. After conditioning on these covariates, mining load appears more comparable outside the SDT window and diverges primarily within it (Figure~\ref{fig2}B--C; see Methods). We therefore estimate the mining-load response associated with the SDT window while conditioning on growth covariates, weather, calendar fixed effects, and load-zone fixed effects.

Within this framework, we model wholesale-price responsiveness using a reduced-translog specification \cite{christensen1973transcendental}. This flexible demand specification allows the electricity-price response of Bitcoin-mining load to vary with hashprice, consistent with the descriptive patterns in Extended Data Figure~\ref{edf:elec-price-hashprice}. Our main specification estimates a common price-response relationship across the three load zones. We also estimate zone-specific specifications to assess whether the response is present within individual zones.

Second, we examine whether mining load responds to the expected opportunity cost created by coincident-peak charges. This opportunity cost is not directly observed, so we proxy for it using a near-peak risk index, denoted NP-risk (see Methods). Higher NP-risk indicates greater proximity to the monthly system peak and therefore a higher expected opportunity cost for consuming electricity. We restrict this analysis to the SDT window, where coincident-peak incentives are concentrated. We also re-estimate the model using an alternative near-peak index to assess whether the results are sensitive to the proxy construction.

\section{Hashprice moderates Bitcoin miners' responsiveness to wholesale electricity prices}\label{sec2}

\begin{figure*}[!htbp]
\centering
\makebox[\textwidth][c]{
\includegraphics[width=1.4\textwidth]{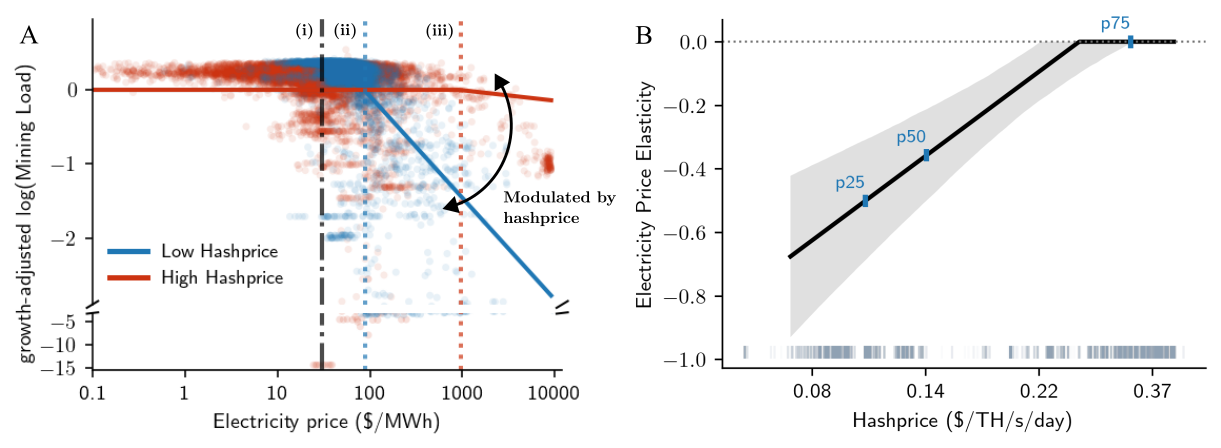}
}
\caption{Threshold-like wholesale-electricity-price responsiveness of Bitcoin-mining load.
(a) Growth-adjusted log mining load plotted against electricity price. Points show hourly observations split into low-hashprice (blue) and high-hashprice (red) groups. Solid lines show model-implied price--response profiles from the matched IV reduced-translog specification in Table~\ref{tab:flexible_load_elasticity}, column~(3). Profiles are evaluated separately for the two hashprice groups, with other covariates held at their observed values. The dash-dotted black line (i) marks a representative contracted power cost of about \$30/MWh reported by large Texas miners \cite{DeRoche2025_Earthjustice}. The dotted vertical lines (ii) and (iii) mark where the fitted low- and high-hashprice profiles cross zero on the growth-adjusted log-load scale. These crossings indicate illustrative model-implied thresholds between the near-capacity region and the curtailment region.
(b) Model-implied electricity-price elasticity at different hashprice levels. Estimates are from the same matched IV reduced-translog specification. The black line shows the point estimate, and the shaded region shows the 95\% confidence interval. The labels p25, p50, and p75 mark the weighted 25th, 50th, and 75th percentiles of hashprice in the matched sample. The rug plot shows the corresponding weighted hashprice distribution.}\label{fig3}
\end{figure*}

Our preferred estimates imply a threshold-like response of aggregate Bitcoin-mining load to wholesale electricity prices (Figure~\ref{fig3}). Although we do not observe individual device-level operations, the fitted aggregate response suggests that mining load reflects heterogeneous devices operating around different breakeven points. Two features of the fitted response support this interpretation. First, at a given hashprice, mining load remains near its effective capacity ceiling when wholesale prices are low but begins to fall once prices rise sufficiently. This implied curtailment threshold shifts to higher wholesale prices when hashprice is higher, consistent with stronger revenue conditions allowing marginal devices to remain profitable at higher electricity costs. Second, within the responsive portion of the demand curve, aggregate load continues to decline as electricity prices rise, suggesting that an increasing share of mining capacity curtails as operating margins deteriorate. Higher hashprice also flattens this responsive region. Within this region, the model-implied electricity-price elasticity is about $-0.5$ at the 25th percentile of hashprice, about $-0.3$ at the median, and close to zero at the 75th percentile.

Table~\ref{tab:flexible_load_elasticity} shows how this pattern emerges across specifications. Column~(1) estimates the broad effect associated with the SDT window, controlling for the growth covariate, weather controls, calendar fixed effects, and load-zone fixed effects. Mining load falls by about 9\% within the SDT window in 2021, with an additional reduction in 2022 that implies total SDT curtailment of roughly 39\%. These estimates are robust to alternative definitions of the SDT window and to an alternative growth covariate (Supplementary Tables~5,~7). Columns~(2) and~(3) add the reduced-translog block, which includes electricity price, hashprice, and their interaction. We treat these terms as endogenous and estimate the specifications using the instrumental-variables (IV) strategy described in the Methods. In the full-sample IV specification in column~(2), the implied electricity-price elasticity, evaluated at \$75/MWh and \$0.15/TH/s/day, is about $-0.12$. The negative electricity-price coefficient and positive electricity-price--hashprice interaction indicate that mining load declines as electricity prices rise, but this response weakens when hashprice is higher. A Wald test rejects the restriction that the electricity-price--hashprice interaction can be omitted ($\chi^2 = 25.34$, $p < 0.001$). However, excluding low-electricity-price observations substantially increases the magnitude of the interaction term (Supplementary Table~10).

\begin{table}[!h]
\caption{Elasticity estimation of Bitcoin mining load with respect to electricity prices and hashprice.}
\label{tab:flexible_load_elasticity}
\begin{tabular}{@{}lllll@{}}
\toprule
 &  & (1) FE-OLS (Baseline) & (2) Full sample & (3) Matched sample \\
 &  &  & ~~~~~IV-2SLS & ~~~~~IV-2SLS \\
\midrule
Dependent variable: & & \multicolumn{3}{c}{$\log(\text{Mining Load})$} \\
\midrule

SDT &  & -0.0921$^{***}$ & -0.1747$^{***}$ & -0.0194 \\
 &  & (0.0279) & (0.0362) & (0.0630) \\[3pt]

SDT $\times$ 2022 &  & -0.4050$^{***}$ & -0.1447$^{**}$ & -0.2623$^{***}$ \\
 &  & (0.0533) & (0.0686) & (0.0649) \\[3pt]

$\log(\text{electricity price})$ &  & --- & -0.1154$^{***}$ & -0.3050$^{***}$ \\
 &  &  & (0.0272) & (0.0636) \\[3pt]

$\log(\text{hashprice})$ &  & --- & 0.2010$^{***}$ & 0.3030$^{**}$ \\
 &  &  & (0.0475) & (0.1269) \\[3pt]

$\log(\text{electricity price}) \times$ &  & --- & 0.2359$^{***}$ & 0.5311$^{***}$ \\
~~~~~~~~~~~~~~~$\log(\text{hashprice})$ &  &  & (0.0469) & (0.1099) \\[3pt]

$\log(\text{growth})$ &  & 0.9906$^{***}$ & 0.9835$^{***}$ & 0.9350$^{***}$ \\
 &  & (0.0299) & (0.0318) & (0.1218) \\[3pt]

temperature &  & 0.0185$^{***}$ & 0.0204$^{***}$ & 0.0431$^{***}$ \\
 &  & (0.0048) & (0.0051) & (0.0158) \\[3pt]

temperature$^{2}$ &  & -0.0001$^{***}$ & -0.0001$^{***}$ & -0.0003$^{***}$ \\
 &  & (0.0000) & (0.0000) & (0.0001) \\[3pt]

Controls and FE &  & Yes & Yes & Yes \\[3pt]

\midrule
Observations &  & 29{,}635 & 29{,}635 & 8{,}637 \\
Covariance Type &  & Clustered & Clustered & Clustered \\[3pt]
$R^2$ (Within) &  & 0.4687 & --- & --- \\
$R^2$ &  & 0.6540 & 0.6562 & 0.6675 \\[3pt]
First-stage Partial $R^2$ &  & --- & 0.18--0.93 & 0.18--0.80 \\
First-stage Partial $F$ &  & --- & 642--20{,}371 & 77--1{,}407 \\
\bottomrule
\end{tabular}

\footnotetext{\textit{Notes:} Cluster-robust standard errors, clustered by date, are reported in parentheses. Asterisks denote statistical significance at the 1\% (***), 5\% (**), and 10\% (*) levels. The dependent variable is $\log(\text{Mining Load})$. All specifications include the growth covariate, 9-hour-lagged ambient temperature and its square, calendar fixed effects, and load-zone fixed effects. First-stage statistics report the range across endogenous regressors. 

Column~(1) reports the baseline fixed-effects OLS specification for SDT effect estimation. Columns~(2) and~(3) incorporate two-stage least-squares (2SLS) estimates for the reduced-translog specification, where $\log(\text{electricity price})$, $\log(\text{hashprice})$, and $\log(\text{electricity price}) \times \log(\text{hashprice})$ are treated as endogenous. The excluded instruments are realized Texas power-grid wind generation, logged Bitcoin price, and their interaction. Price variables are centered at $\$75$/MWh for electricity price and $\$0.15$/TH/s/day for hashprice, so lower-order coefficients in the translog specification are interpreted at those reference points.

Column~(3) is estimated on the matched sample using CEM weights. Matching is implemented separately by year using Texas power-grid day-ahead load forecasts and renewable-generation forecasts. Across years, the matched sample retains 29.1\% of observations, including 91.1\% of treated observations and 20.5\% of control observations. Matching reduces overall $L_1$ imbalance from 0.847 to approximately zero in 2021 and from 0.876 to approximately zero in 2022.}

\end{table}

Our preferred specification in column~(3) re-estimates the reduced-translog model on a matched sample constructed using coarsened exact matching (CEM) \cite{iacus2012causal} on system-wide demand and renewable-generation forecasts. Matching improves comparability between SDT and non-SDT observations by ensuring that they occur under similar forecasted grid conditions (Supplementary Figure~7). However, matching also reduces the influence of low-load, high-renewable states in the reduced-translog model, in which mining load is often near its effective capacity ceiling (Supplementary Figure~8). The matched-sample estimate implies an elasticity of about $-0.30$ at $\$75$/MWh and $\$0.15$/TH/s/day. By contrast, the corresponding elasticity is about $-0.05$ in the excluded observations. The matched-sample estimate also remains stable when low-price observations are removed (Supplementary Tables~11--12). We therefore use the matched-sample IV reduced-translog model as our preferred specification. Because the reduced-translog specification is smooth and unconstrained, it can extrapolate outside the physically and economically meaningful region for Bitcoin miners. We therefore interpret the fitted response only within the region defined by non-positive electricity-price elasticity, non-negative hashprice elasticity, and fitted growth-adjusted log mining load no greater than zero. Outside this region, we interpret mining load as operating near an effective capacity ceiling (Supplementary Note~2; Supplementary Figure~9). Figure~\ref{fig3} illustrates the fitted demand curves implied by this interpretation.

Re-estimating the model separately by load zone yields a similar wholesale-price response despite smaller samples. The electricity-price coefficients remain negative, and joint Wald tests reject the null that the translog price terms are jointly zero in each zone (Supplementary Table~13). The magnitudes vary across zones, consistent with regional differences in mining fleet composition. Our preferred wholesale-electricity-price measure is the average of day-ahead and real-time prices; results are similar when using day-ahead or real-time prices separately (Supplementary Table~9).

\section{Mining load responds to incentives created by coincident-peak-based transmission charges}\label{sec3}

The wholesale-price results show that aggregate Bitcoin-mining load decreases when wholesale electricity prices rise, and this response weakens when expected mining revenue is higher. Coincident-peak-based transmission charges create a second electricity-sector cost channel. Unlike wholesale electricity prices, the cost created by coincident-peak charges is not observed as a contemporaneous price. We therefore proxy for this cost using near-peak risk, which measures how close an interval is to becoming a coincident peak. Higher near-peak risk implies a higher expected opportunity cost of electricity consumption. If miners internalize this opportunity cost, aggregate mining load should fall as near-peak risk rises, and this response should be weaker when hashprice is higher.

\begin{table}[!htbp]
\caption{Bitcoin-mining load response to near-peak risk, wholesale electricity prices, and hashprice.}
\label{tab:capacity_elasticity}
\begin{tabular}{@{}lccc@{}}
\toprule
 & (4) NP-risk block & (5) Electricity price  & (6) NP-risk block ~~~~~~~~~~ \\
 &   & block~~~~~~~~        & + Electricity price block \\
\midrule
Dependent variable: & \multicolumn{3}{c}{$\log(\text{Mining Load})$} \\
\midrule

NP-risk   & -1.1831$^{***}$ & --- & -0.8313$^{***}$ \\
                & (0.1784)        &     & (0.1579)        \\[3pt]

NP-risk   & 1.1389$^{***}$ & --- & 0.6018$^{***}$ \\
~~~~~~~~~~~~$\times \log(\text{hashprice})$
                & (0.2265)       &     & (0.2123)        \\[3pt]

$\log(\text{electricity price})$    & --- & -0.3231$^{***}$ & -0.2349$^{***}$ \\
                                    &     & (0.0554)        & (0.0568)        \\[3pt]

$\log(\text{electricity price})$    & --- & 0.6530$^{***}$ & 0.6210$^{***}$ \\
~~~~~~~~~~~$\times \log(\text{hashprice})$
                                    &     & (0.0845)       & (0.0852)        \\[3pt]

$\log(\text{hashprice})$    & -0.8109$^{***}$ & -0.5553$^{***}$ & -0.7706$^{***}$ \\
                            & (0.2002)        & (0.2015)        & (0.2011)        \\[3pt]

$\log(\text{growth})$       & 1.1092$^{***}$ & 1.2015$^{***}$ & 1.1575$^{***}$ \\
                            & (0.1034)       & (0.1047)       & (0.1048)        \\[3pt]

temperature                 & -0.0523$^{**}$ & -0.0311 & -0.0433$^{**}$ \\
                            & (0.0235)       & (0.0218) & (0.0218)       \\[3pt]

temperature$^2$             & 0.0004$^{**}$ & 0.0001 & 0.0003$^{**}$ \\
                            & (0.0002)      & (0.0002) & (0.0002)      \\[3pt]

Controls and FE             & Yes & Yes & Yes \\
\midrule
Observations                & 3{,}630 & 3{,}630 & 3{,}630 \\
Covariance type             & Clustered & Clustered & Clustered \\[3pt]
$R^2$ (overall)             & 0.3583 & 0.4004 & 0.4160 \\[3pt]
First-stage partial $R^2$   & 0.47--0.85 & 0.32--0.85 & 0.32--0.85 \\
First-stage partial $F$     & 324--2{,}947 & 165--2{,}947 & 165--2{,}947 \\
\bottomrule
\end{tabular}

\footnotetext{\textit{Notes:} Cluster-robust standard errors, clustered by date, are reported in parentheses. All columns use hourly summer--daytime observations. Asterisks denote statistical significance at the 1\% (***), 5\% (**), and 10\% (*) levels. The dependent variable is $\log(\text{Mining Load})$. All specifications include the growth covariate, 9-hour-lagged ambient temperature and its square, calendar fixed effects, and load-zone fixed effects. First-stage statistics report the range across endogenous regressors.

Column~(4) includes the NP-risk--hashprice block, column~(5) includes the electricity-price--hashprice block, and column~(6) includes both blocks jointly. The endogenous variables are the included NP-risk block, electricity-price block, and logged hashprice, as applicable in each specification, estimated using 2SLS. The excluded instruments are held fixed across columns and consist of Houston cooling degree days, Texas power-grid wind generation, logged Bitcoin price, and the corresponding interaction instruments. Variables are centered at $0.50$ for NP-risk, $\$75$/MWh for electricity price, and $\$0.15$/TH/s/day for hashprice, so lower-order coefficients in the specifications are interpreted at those reference points.}

\end{table}

We test this prediction by estimating whether aggregate mining load declines with NP-risk and whether this relationship is moderated by hashprice. Because these opportunity costs are relevant primarily during coincident-peak determination periods, we restrict the analysis to SDT observations. Because NP-risk may be correlated with contemporaneous wholesale electricity prices, we compare specifications that include the NP-risk--hashprice block, the electricity-price--hashprice block, and both blocks jointly. The joint specification, reported in column~(6) of Table~\ref{tab:capacity_elasticity}, is our preferred specification because it allows us to assess whether the relationship between mining load and near-peak risk persists after accounting for the contemporaneous wholesale-price signal. All specifications include the growth covariate, weather, calendar fixed effects, and load-zone fixed effects. We treat the included NP-risk, wholesale-price, hashprice, and interaction terms as endogenous and estimate the specifications using the IV strategy described in the Methods.

\begin{figure}[!htbp]
\centering
\includegraphics[width=0.77\textwidth]{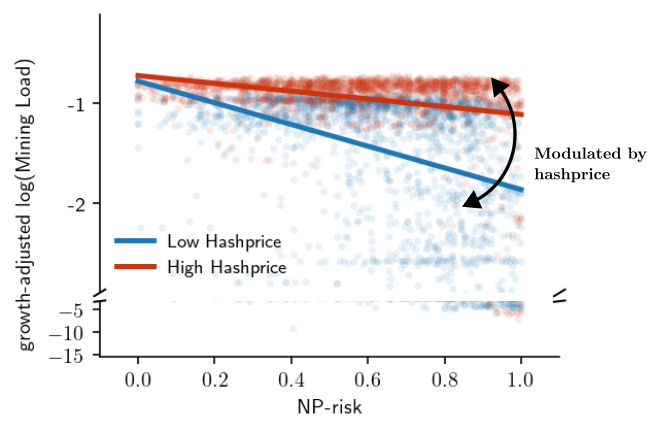}
\caption{Hashprice moderates the relationship between near-peak risk and Bitcoin-mining load.
Growth-adjusted log Bitcoin-mining load is plotted against the near-peak risk index, NP-risk. Points represent hourly SDT observations split into low- and high-hashprice groups. Solid lines show model-implied responses from the specification in Table~\ref{tab:capacity_elasticity}, column~(6). Higher NP-risk indicates a greater likelihood of contributing to future transmission charges.}\label{fig4}
\end{figure}

The estimates support the predicted state-dependent response. In the NP-risk-only specification, higher NP-risk is associated with lower mining load, while the positive NP-risk--hashprice interaction indicates that this relationship weakens when hashprice is higher. When the NP-risk and electricity-price blocks are included jointly, the NP-risk coefficients attenuate but remain statistically and economically meaningful. A joint Wald test rejects the null that the NP-risk terms are jointly zero ($\chi^2 = 28.18$, $p<0.001$). We also re-estimate these specifications using an alternative NP-risk measure; the estimated coefficients preserve the same sign pattern (Methods; Supplementary Table~16). These results suggest that mining load responds to coincident-peak incentives in a way that is not solely explained by contemporaneous wholesale electricity prices. Figure~\ref{fig4} illustrates the implied response, mirroring the state dependence observed in the wholesale-price response. Finally, because coincident-peak incentives should be most relevant during the SDT window, we estimate placebo specifications for summer nights and non-summer windows. The SDT specification is the only one that exhibits the expected sign pattern (Supplementary Table~17).

\section{Discussion}\label{sec4}

Aggregate Bitcoin-mining load is flexible, but not in a fixed or unconditional way. Its responsiveness to electricity-sector costs depends on the revenue state of mining. We find that mining load declines when electricity-sector costs rise, and this response weakens when hashprice is higher. This state dependence appears across two distinct cost channels, contemporaneous wholesale electricity prices and expected opportunity costs created by coincident-peak-based transmission charges. The wholesale-price response provides the clearest evidence of the mechanism underlying Bitcoin miners' demand response: aggregate mining load remains near its effective capacity ceiling at low prices, then begins to fall beyond an implied curtailment threshold. Higher hashprice shifts this threshold toward higher electricity prices. This behavior is consistent with heterogeneous mining devices operating around different breakeven points. As electricity-sector costs rise, devices with lower breakeven prices are curtailed first, generating the aggregate threshold-like response. Higher expected mining revenue raises these breakeven prices, allowing mining load to remain online at electricity prices that would otherwise induce curtailment.

This state dependence matters for power systems. Bitcoin-mining facilities may appear to offer a large source of demand response, but the flexibility available to the grid depends partly on revenue conditions in the crypto-financial sector. Treating such load as a stable flexibility resource may therefore overstate available demand response during periods when mining revenues are high. This interpretation is subject to several limitations. Mining-load data are aggregated to load zones, coverage differs across zones over time, hashprice is observed daily, whereas electricity prices are observed hourly, and we do not observe device-level operations, contractual exposure, or firm-specific curtailment strategies. Nevertheless, the consistency of the estimated responses across wholesale-price and coincident-peak cost channels supports the interpretation that Bitcoin-mining flexibility is economically state dependent. This state dependence may not be unique to Bitcoin mining. Similar patterns may arise in other emerging flexible-load sectors with sector-specific revenue couplings, including hydrogen electrolysis \cite{ruhnau2022flexible}, energy-intensive AI data centers \cite{colangelo2026ai}, and other electricity-intensive processes often assumed to be flexible. Estimating demand-side flexibility for planning, operations, and market design therefore requires identifying not only the electricity-sector signals that induce curtailment, but also the external revenue conditions that determine whether curtailment is economically attractive.

\section{Methods}\label{sec11}

\subsection*{Data}
We construct an hourly panel linking Bitcoin-mining-related electricity-consumption data to mining-revenue conditions, ERCOT wholesale electricity prices, ERCOT grid conditions, and local weather. ERCOT operates most of the Texas electricity grid. In the main text, we use ``Texas power market'' and ``Texas power grid'' as shorthand for ERCOT; in the Methods, we use ``ERCOT'' to match the terminology used in data sources and market rules.

ERCOT provided confidential hourly observations of Large Flexible Load (LFL) power consumption aggregated to the load-zone level under a data-use agreement. The sample covers three ERCOT load zones: West, from January 1, 2021, through July 24, 2022; North, from March 2, 2022, through July 24, 2022; and South, from June 15, 2021, through December 12, 2022. We use load-zone LFL consumption as a proxy for Bitcoin-mining-related load. Because the LFL category may include some flexible loads that are not Bitcoin miners, we interpret our estimates as responses of ERCOT LFL consumption associated with Bitcoin-mining activity. This interpretation is supported by ERCOT documentation linking LFL forecasts to crypto-mining facilities and Bitcoin-market conditions \cite{ERCOT_MORA_May2025}.

LFL consumption is measured in MW and can equal zero during curtailment episodes. To retain zero-load observations in logarithmic specifications, we add an offset of 0.0001 MW before taking logs. For notational simplicity, we refer to the resulting transformed outcome for load zone $z$ and hour $t$ as $\log(\text{Mining Load}_{zt})$ throughout.

We measure mining-revenue conditions using hashprice, defined as expected U.S.-dollar-denominated mining revenue per unit of computational power per unit time \cite{LuxorHashprice2025, neumueller2025cambridge}. We use the daily hashprice series published by Luxor and obtain both hashprice and Bitcoin price data from \url{https://hashrateindex.com/}. These daily series are merged into the hourly panel by calendar date. The hashprice series is shown in Supplementary Figure~6.

Wholesale electricity prices are measured using hourly ERCOT day-ahead market (DAM) and real-time market (RTM) prices for each load zone. For the main elasticity specifications, we summarize wholesale price conditions using the average of DAM and RTM prices:
\begin{equation}
P_{zt}
= \frac{\text{DAM Price}_{zt} + \text{RTM Price}_{zt}}{2}.
\end{equation}
We log-transform this variable in the empirical specifications. Hours in which the averaged electricity price is non-positive are excluded because the logarithm is undefined for these observations.

ERCOT grid conditions are measured using public day-ahead forecasts of ERCOT-wide demand, wind generation, and solar generation. For each operating day $d$, we use forecasts posted on day $d-1$ before the 10{:}00 cutoff for day-ahead market participation \cite{ERCOT_Nodal_Protocols_2025}. Specifically, we use the system load forecast posted around 09{:}30 and the wind and solar generation forecasts posted around 09{:}55. These forecasts were publicly available before the day-ahead market participation cutoff and therefore represent information that could have been available to miners when forming expectations about next-day operating conditions (Supplementary Figure~1). We sum the wind and solar forecasts to construct a renewable-generation forecast. We also collect realized ERCOT demand and realized ERCOT wind generation. Historical ERCOT demand from 2015 through 2020 is collected separately to define the summer--daytime (SDT) window described below.

To capture local ambient conditions relevant to mining operations, we assign a representative dry-bulb temperature series to each ERCOT load zone. Representative locations are chosen based on the approximate locations of large-scale Bitcoin-mining facilities shown in Supplementary Figure~2. Hourly dry-bulb temperature data are obtained from \url{https://meteostat.net}. We also collect Houston dry-bulb temperature and construct daily cooling degree days (CDD) relative to a base temperature of 65$^\circ$F. Daily CDD values are merged into the hourly panel by calendar date.

To proxy for effective installed mining capacity, we construct a growth covariate using only prior mining-load observations outside the SDT window:
\begin{equation} 
\text{growth}_{zt} =  \max_{\substack{s < t, \ s \notin \text{SDT}}} \text{Mining Load}_{zs}. 
\end{equation}
This measure depends only on mining load observed before hour $t$ and outside the SDT window, so it does not mechanically incorporate contemporaneous SDT curtailment. In the estimation sample, $\text{growth}_{zt}$ is strictly positive, allowing us to include $\log(\text{growth}_{zt})$ in the empirical specifications. The resulting growth covariate for each load zone is shown in Figure~\ref{fig2}B. As a robustness check, we also construct an alternative growth covariate using a piecewise linear fit to non-SDT mining load.

All timestamps are converted to a fixed Central Standard Time (CST) convention before merging. This avoids discontinuities associated with daylight saving time transitions and ensures that calendar variables are defined consistently across the panel. Hour-of-day, day-of-week, month, and year indicators are constructed using this fixed-time convention. Hours with missing observations are excluded from the analysis. Because LFL data become available on different dates across load zones, the panel is unbalanced. All specifications use the available zone-hour observations. Descriptive statistics for the analysis sample are reported in Supplementary Table~1.

\subsection*{SDT window}

Under Public Utility Commission of Texas rules, coincident peaks are determined separately for each summer month, June through September, based on system peak demand \cite{PUCT1999_25_192}. We define the summer--daytime (SDT) window using historical ERCOT-wide demand from 2015--2020 and ERCOT estimates of the maximum load actively pursuing reduction during coincident-peak intervals (Supplementary Table~2). This pre-sample procedure identifies hours in which reported responsive load could have affected which interval set the coincident peak.

For year $y$, summer month $m$, and hour $t$, let $\text{Demand}_{ymt}$ denote realized ERCOT system demand, $\text{Curtailable\_Load}_{y}$ denote ERCOT's reported maximum responsive load during coincident-peak intervals, and $\text{Peak\_Demand}_{ym}$ denote the realized monthly system peak. We define near-peak hours as
\begin{equation}
\begin{split}
H^{\text{near-peak}}_{ym}
&= \bigl\{ t \in (y,m) :
\ \text{Demand}_{ymt} + \text{Curtailable\_Load}_y \\
&\hphantom{= \bigl\{ t \in (y,m) :} \ge \text{Peak\_Demand}_{ym} \bigr\}.
\end{split}
\end{equation}

Pooling near-peak hours across 2015--2020 shows that they are concentrated between 12{:}00 and 19{:}00 using the fixed CST convention and can occur on both weekdays and weekends (Figure~\ref{fig2}A). We therefore define the SDT window as June through September, 12{:}00--19{:}00 CST. Because this definition uses only pre-sample ERCOT demand and published curtailment estimates, it is determined independently of the mining-load outcomes analyzed in the estimation sample. This timing is consistent with documented coincident-peak intervals during 2001--2014 and with inferred miner-curtailment patterns in 2021--2022 (Supplementary Figures~3--5).

\subsection*{Difference-in-differences specification}
We use the SDT window to estimate broad curtailment responses associated with exposure to coincident-peak incentives. SDT window is defined as June--September, 12{:}00--19{:}00. We treat SDT hours as exposed observations and non-SDT hours as comparison observations.

The estimating equation is
\begin{equation}
\label{eq:did_main}
\begin{aligned}
\log(\text{Mining Load}_{zt})
= {} & \alpha_z + \gamma_t
+ \beta_1 \,\text{SDT}_{t}
+ \beta_2 \,\text{SDT}_{t} \times \mathbf{1}\{t \in 2022\} \\
& {} +  \theta\,\log\left(\text{growth}_{zt}\right)
+ f(\text{Weather}_{zt})
+ \epsilon_{zt},
\end{aligned}
\end{equation}
where $\alpha_z$ are load-zone fixed effects and $\gamma_t$ includes fixed effects for hour of day, day of week, month, and year. The indicator $\mathbf{1}\{t \in 2022\}$ allows the SDT-window response to differ between 2021 and 2022. The variable $\text{growth}_{zt}$ is the preferred capacity-growth covariate. The weather control function $f(\text{Weather}_{zt})$ includes 9-hour-lagged temperature and its square. Standard errors are clustered by date.

We interpret the SDT coefficients as intention-to-treat responses. This interpretation relies on three considerations. First, the timing of SDT exposure must be plausibly exogenous to realized mining-load outcomes. The summer months are institutionally designated, and the intraday SDT window is defined using ERCOT system-load patterns from 2015--2020, before the mining-load sample.

Second, conditional on fixed effects, weather controls, and capacity-growth adjustment, non-SDT observations must provide a valid counterfactual for SDT observations absent SDT-related coincident-peak incentives. In difference-in-differences terms, SDT and non-SDT mining loads should have followed parallel trends in the absence of these incentives. We address the capacity growth-concern using the preferred growth covariate. The growth-adjusted patterns in Figure~\ref{fig2}C are consistent with this parallel-trends requirement.

Third, exposed and comparison observations should have common support in system conditions relevant to miners' operating conditions. To assess this requirement, we also estimate specifications on a matched sample constructed using ERCOT day-ahead load forecasts and renewable-generation forecasts, which were publicly available before the day-ahead market participation cutoff. Matching is implemented separately by calendar year and substantially reduces imbalance in these forecast covariates; the $L_1$ imbalance measure falls to approximately zero in both years (Supplementary Figure~7).

In specifications estimated on the matched sample, month-by-year fixed effects are omitted because they make identification rely on relatively thin within-month-year SDT versus non-SDT contrasts. Specifications including month-by-year fixed effects are reported in Supplementary Table~6. We use lagged temperature to allow observed mining load to reflect delayed responses to ambient conditions, such as thermal inertia in mining equipment or cooling systems. The estimates are robust to alternative temperature lags (Supplementary Table~8).

\subsection*{IV reduced-translog specification}

We estimate a reduced-translog specification using two-stage least squares (2SLS) to describe how Bitcoin-mining load varies with wholesale electricity prices and hashprice. The specification allows the electricity-price response to depend on mining-revenue conditions. The SDT block is included in the control vector.

The second-stage equation is
\begin{equation}
\label{eq:second_stage_translog}
\begin{aligned}
\log(\text{Mining Load}_{zt})
= {} & \alpha_z + \gamma_t
+ \beta_p \log P_{zt}
+ \beta_h \log H_t
+ \beta_{ph}\big(\log P_{zt}\,\log H_t\big) \\
& {} + \delta^\top X_{zt}
+ u_{zt},
\end{aligned}
\end{equation}
where $P_{zt}$ is the average wholesale electricity price in load zone $z$ and hour $t$, and $H_t$ is hashprice. The terms $\alpha_z$ are load-zone fixed effects, and $\gamma_t$ includes fixed effects for hour of day, day of week, month, and year. The control vector $X_{zt}$ includes $\text{SDT}_{t}$, $\text{SDT}_{t}\times \mathbf{1}\{t\in 2022\}$, $\log(\text{growth}_{zt})$, 9-hour-lagged temperature, and lagged temperature squared. Standard errors are clustered by date.

We treat $\log P_{zt}$, $\log H_t$, and their interaction as endogenous. The excluded instruments are ERCOT-wide realized wind generation, logged Bitcoin price, and their interaction. The identifying assumption is that, conditional on controls and fixed effects, these instruments affect mining load only through wholesale electricity prices, hashprice, and their interaction. First-stage and endogeneity diagnostics are reported in Supplementary Note~1 and Supplementary Tables~3--4.

For each endogenous regressor $Z_{zt}\in\{\log P_{zt},\log H_t,\log P_{zt}\log H_t\}$, the first-stage equation is
\begin{equation}
\label{eq:first_stage_generic}
\begin{aligned}
Z_{zt}
= {} & \alpha_z + \gamma_t
+ \pi_1\,\text{Wind}_t
+ \pi_2\,\log(\text{BTC price}_t) \\
& {} ~~~~~~~~~~~~~~~~~~~~~~~~~~~~~~~~~+ \pi_3\,\big(\text{Wind}_t \times \log(\text{BTC price}_t)\big) + \rho^\top X_{zt}
+ \eta_{zt}.
\end{aligned}
\end{equation}

The implied electricity-price elasticity is $\varepsilon_p(P_{zt},H_t)=\beta_p+\beta_{ph}\log H_t$, and the implied hashprice elasticity is $\varepsilon_h(P_{zt},H_t)=\beta_h+\beta_{ph}\log P_{zt}$. Our preferred reduced-translog specification is estimated on the matched sample described above.

In estimation, the logged price variables are centered at $\$75$/MWh for wholesale electricity prices and $\$0.15$/TH/s/day for hashprice:
\begin{equation}
\widetilde{\log P}_{zt} = \log(P_{zt}) - \log(75),
\qquad
\widetilde{\log H}_{t} = \log(H_t) - \log(0.15).
\end{equation}
To keep the notation compact, equation~\eqref{eq:second_stage_translog} writes these centered variables as $\log P_{zt}$ and $\log H_t$. With this centering, the lower-order coefficients $\beta_p$ and $\beta_h$ are interpreted as the electricity-price and hashprice elasticities, respectively, at $\$75$/MWh and $\$0.15$/TH/s/day.

\subsection*{Near-peak risk specification}

We restrict this analysis to SDT hours and construct a near-peak risk index, denoted NP-risk, using realized ERCOT-wide system load. The index proxies for the expected opportunity cost created by coincident-peak charges. We normalize system demand within each year--month because coincident peaks are determined separately for each summer month. Let $t$ denote an SDT hour in summer month $m$ and year $y$, and let $\mathcal{T}^{\text{SDT}}_{ym}$ denote the set of SDT hours in the same year--month. We define
\begin{equation}
\label{eq:np_index}
\text{NP-risk}_{t} = \frac{\text{Demand}_{ymt} - \min_{t' \in \mathcal{T}^{\text{SDT}}_{ym}} \text{Demand}_{ymt'}}
     {\max_{t' \in \mathcal{T}^{\text{SDT}}_{ym}} \text{Demand}_{ymt'} - \min_{t' \in \mathcal{T}^{\text{SDT}}_{ym}} \text{Demand}_{ymt'}}.
\end{equation}
This transformation maps realized ERCOT-wide system load into the unit interval within each year--month, with larger values indicating hours closer to the monthly peak.

Within SDT hours, we estimate
\begin{equation}
\label{eq:iv_main}
\begin{aligned}
\log(\text{Mining Load}_{zt})
= {} & \alpha_z + \gamma_t \\
& {} + \beta_R\,\text{NP-risk}_{t}
      + \beta_{RH}\,\text{NP-risk}_{t}\,\log H_t \\
& {} + \beta_P\,\log P_{zt}
      + \beta_{PH}\,\log P_{zt}\,\log H_t \\
& {} + \beta_H\,\log H_t
      + \theta^\top X_{zt}
      + \nu_{zt},
\end{aligned}
\end{equation}
where $\alpha_z$ are load-zone fixed effects and $\gamma_t$ includes fixed effects for hour of day, day of week, and month-by-year. The control vector $X_{zt}$ includes $\log(\text{growth}_{zt})$, 9-hour-lagged temperature, and lagged temperature squared. Standard errors are clustered by date. Month-by-year fixed effects absorb year-month-specific level shifts in summer operating conditions. This specification is motivated by Extended Data Figure~\ref{edf:np-risk-hashprice-ercot}, which suggests that, after conditioning on the growth covariate, remaining differences across months and years primarily reflect level shifts.

We estimate the model using 2SLS. We treat the NP-risk block, $\{\text{NP-risk}_{t},\text{NP-risk}_{t}\log H_t\}$, the electricity-price block, $\{\log P_{zt},\log P_{zt}\log H_t\}$, and $\log H_t$ as endogenous. The excluded instruments are Houston cooling degree days, realized ERCOT-wide wind generation, logged Bitcoin price, and interaction instruments formed from these variables corresponding to the endogenous interaction terms. The identifying assumption is that, conditional on controls and fixed effects, these instruments affect mining load only through near-peak risk, wholesale electricity prices, hashprice, and their interactions. First-stage and endogeneity diagnostics are reported in Supplementary Tables~14--15.

Using the same outcome, controls, fixed effects, and excluded instrument set, we also estimate two restricted specifications: one excluding the NP-risk block and one excluding the electricity-price block. We report joint Wald tests for the corresponding coefficient blocks in the unrestricted specification.

\subsection*{Alternative near-peak risk definition}

As a robustness check, we construct an alternative near-peak risk index based on each day's maximum ERCOT system load. This alternative measure varies across days rather than across hours and captures whether a given day is close to the monthly peak-demand day.

For day $d$ in month $m$ and year $y$, let $\text{PeakLoad}_{ymd}$ denote the maximum ERCOT system load on that day, and let $\mathcal{D}_{ym}$ denote all days in the same year--month. We define
\begin{equation}
\label{eq:np_index_daily}
\text{NP-risk}_{d}
=
\frac{\text{PeakLoad}_{ymd} - \min_{d'\in \mathcal{D}_{ym}} \text{PeakLoad}_{ymd'}}
     {\max_{d'\in \mathcal{D}_{ym}} \text{PeakLoad}_{ymd'} - \min_{d'\in \mathcal{D}_{ym}} \text{PeakLoad}_{ymd'}}.
\end{equation}
The resulting value is assigned to each SDT hour on day $d$, and the specification in equation~\eqref{eq:iv_main} is re-estimated using this alternative index.

\begin{extdatafigure}[!htbp]
  \label{edf:elec-price-hashprice}
  \includegraphics[width=\textwidth]{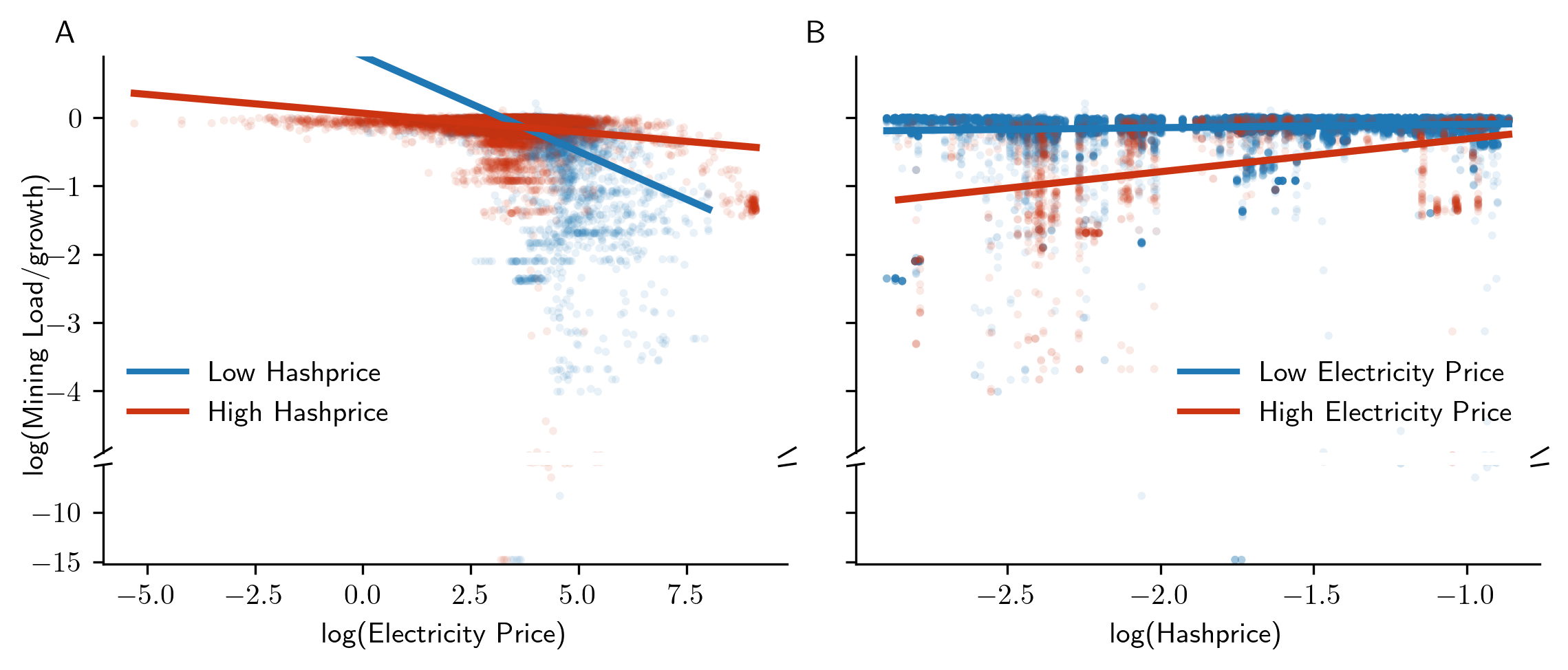}
  \extdatacaption{Growth-adjusted Bitcoin-mining load in relation to logged electricity prices and logged hashprice.
(A) $\log(\text{Mining Load}/\text{growth})$ versus $\log(\text{Electricity Price})$, split by low- and high-hashprice observations.
(B) $\log(\text{Mining Load}/\text{growth})$ versus $\log(\text{Hashprice})$, split by low- and high-electricity-price observations.
Here, ``growth'' is the zone-specific installed-capacity proxy used in the empirical analysis. Points show zone-hour observations, and solid lines show group-specific linear fits. Low-load observations are more frequent at high electricity prices, especially when hashprice is low. Conversely, the relationship between hashprice and growth-adjusted mining load is steeper during high-electricity-price periods. These descriptive patterns are consistent with electricity-price responsiveness varying with hashprice. The broken y-axis preserves extreme low-load observations while keeping the main variation visible.}
\label{edf:elec-price-hashprice}
\end{extdatafigure}

\begin{extdatafigure}[!htbp]
  \label{edf:np-risk-hashprice-ercot}
  \includegraphics[width=\textwidth]{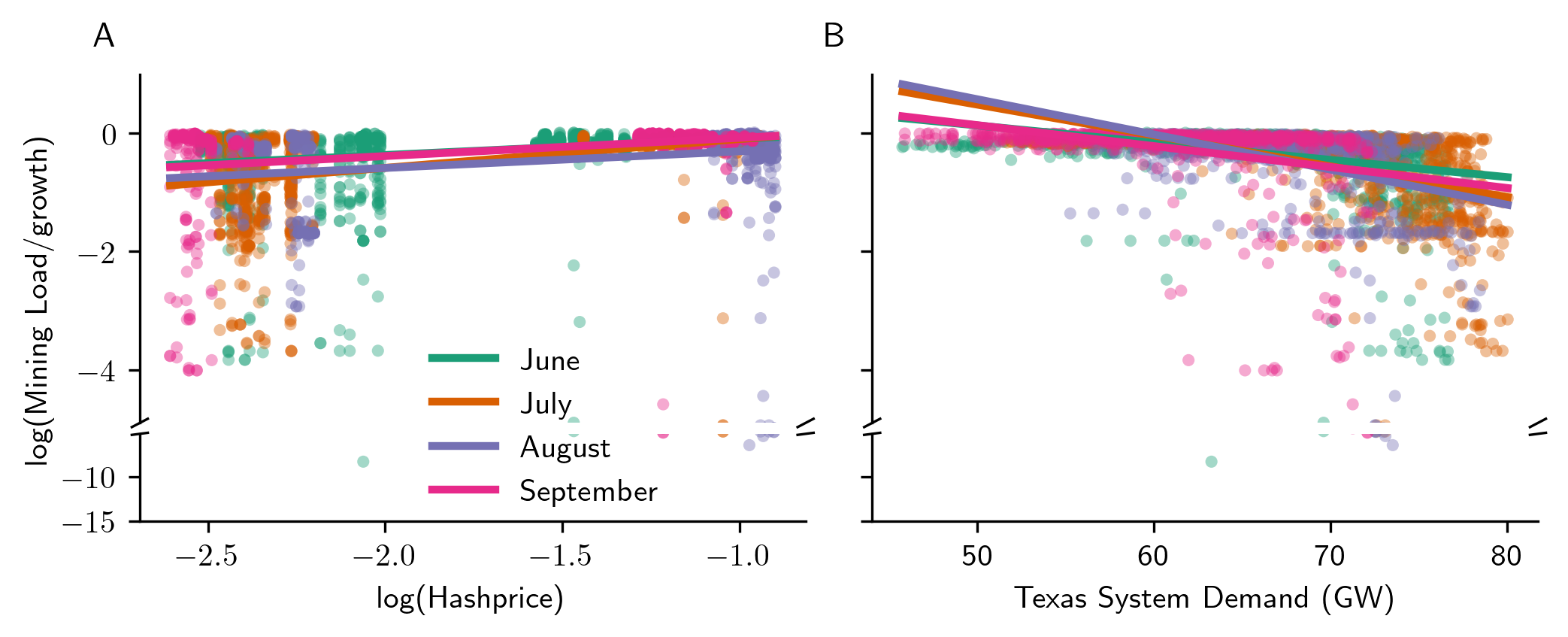}
  \extdatacaption{Growth-adjusted Bitcoin-mining load in relation to logged hashprice and Texas system demand during summer daytime hours.
(A) $\log(\text{Mining Load}/\text{growth})$ versus $\log(\text{Hashprice})$.
(B) $\log(\text{Mining Load}/\text{growth})$ versus Texas system demand.
Here, ``growth'' is the zone-specific installed-capacity proxy used in the empirical analysis. Points show zone-hour observations, with colors indicating summer months from June through September, and solid lines show month-specific linear fits. Growth-adjusted mining load tends to be higher when hashprice is higher and lower when Texas system demand is higher. The fitted relationships are broadly similar across summer months, suggesting that the descriptive patterns are not driven by a single month. The broken y-axis preserves extreme low-load observations while keeping the main variation visible.}
\label{edf:np-risk-hashprice-ercot}
\end{extdatafigure}

\backmatter

\bmhead{Supplementary information}

This manuscript has supplementary information in the accompanying PDF.  

\bmhead{Acknowledgements}
The author gratefully acknowledges thoughtful discussions with Bitcoin miners and ERCOT participants involved in the Blockchain and Energy Research Consortium at Texas A\&M University. The author thanks Le Xie for facilitating access to the proprietary large flexible load power-consumption dataset. The author also thanks Ignacio Aravena and Le Xie for mentorship and funding support. The author used ChatGPT for code-generation assistance and writing support. The author takes full responsibility for the data analysis, and conclusions of this manuscript.

\section*{Declarations}


\begin{itemize}
\item Funding: This work was supported in part by the U.S. Department of Energy (DOE) through the OPEN COG Grid project and in part by the Blockchain and Energy Research Consortium at Texas A\&M University.
\item Conflict of interest/Competing interests: The author declares no competing interests.
\item Ethics approval and consent to participate: Not applicable.
\item Consent for publication: The author has consented to publication of this manuscript.
\item Data availability: Publicly shareable input data are provided in the replication package deposited on Zenodo: \url{https://doi.org/10.5281/zenodo.20272757}. The restricted LFL data cannot be redistributed because they are the property of ERCOT and were accessed under a data-use agreement. These restricted data are merged with public data using timestamps and ERCOT load zone. Access to the restricted LFL data is subject to ERCOT approval.
\item Materials availability: Not applicable.
\item Code availability: The replication package deposited on Zenodo, \url{https://doi.org/10.5281/zenodo.20272757}, provides the code used to generate the figures and tables in the main text and supplementary materials, together with the generated outputs. Some scripts require the restricted ERCOT data and therefore cannot be fully executed using only the public replication package.
\item Author contribution: S.M. conceived the research, developed the empirical framework and identification strategy, conducted the data analysis, and wrote the original draft.
\end{itemize}

\bibliography{sn-bibliography}

\end{document}